\documentclass[prd,aps,twocolumn,amsmath,amssymb,nofootinbib,preprintnumbers]
{revtex4}

\usepackage{graphicx}
\usepackage{dcolumn}
\usepackage{bm}
\usepackage{amsmath}
\usepackage{amsthm}

\usepackage{amsfonts}
\usepackage{euscript,bbm}
\usepackage{ifthen}
\usepackage{psfrag}
\usepackage{slashed}

\def\ls{\mathrel{\lower4pt\vbox{\lineskip=0pt\baselineskip=0pt
           \hbox{$<$}\hbox{$\sim$}}}}
\def\gs{\mathrel{\lower4pt\vbox{\lineskip=0pt\baselineskip=0pt
           \hbox{$>$}\hbox{$\sim$}}}}
\def\drawbox#1#2{\hrule height#2pt

\hbox{\vrule width#2pt height#1pt \kern#1pt
              \vrule width#2pt}
              \hrule height#2pt}

\def\Asym#1#2{\vcenter{\vbox{\drawbox{#1}{#2}
              \kern-#2pt       
              \drawbox{#1}{#2}}}}

\newcommand{\be}{\begin{equation}}
\newcommand{\ee}{\end{equation}}
\newcommand{\bea}{\begin{eqnarray}}
\newcommand{\eea}{\end{eqnarray}}

\newcommand{\neu}[1]{\ensuremath{\tilde{\chi}_{#1}^0}}

\newcommand{\chpm}[1]{\ensuremath{\tilde{\chi}_{#1}^{\pm}}}

\newcommand{\st}{\ensuremath{\tilde{t}}}

\newcommand{\gsim}{\lower.7ex\hbox{$\;\stackrel{\textstyle>}{\sim}\;$}}
\newcommand{\lsim}{\lower.7ex\hbox{$\;\stackrel{\textstyle<}{\sim}\;$}}

\newcommand{\ttbar}{t \bar{t}}

\newcommand{\met} {{E\!\!\!\!/_{\rm T}}}
\newcommand{\HT} {{H_{\rm T}}}
\newcommand{\pT} {{p_{\rm T}}}

\newcommand{ \pythia } {{\tt PYTHIA}}
\newcommand{ \isasugra } {{\tt ISASUGRA}}
\newcommand{ \isajet }    {{\tt ISAJET}}
\newcommand{ \pgs }    {{\tt PGS4}}
\newcommand{ \darksusy }    {{\tt darkSUSY}}

\begin{document}

%
\title{Top Squark Searches Using Dilepton Invariant Mass Distributions and Bino-Higgsino Dark Matter at the LHC}


\author{Bhaskar Dutta$^{1}$}
\author{Teruki Kamon$^{1,2}$}
\author{Nikolay Kolev$^{3}$}
\author{Kuver Sinha$^{1}$}
\author{Kechen Wang$^{1}$}
\author{Sean Wu$^{1}$}

\affiliation{$^{1}$~Mitchell Institute for Fundamental Physics and Astronomy,
Department of Physics and Astronomy, Texas A\&M University, College Station, TX 77843-4242, USA \\
$^{2}$~Department of Physics, Kyungpook National University, Daegu 702-701, South Korea \\
$^{3}$~Department of Physics, University of Regina, SK, S4S 0A2, Canada
}

\begin{abstract}

Pair production of light top squarks at the $8$-TeV LHC can be used to probe the gaugino-Higgsino sector of the Minimal Supersymmetric Standard Model. The case where the lightest neutralino is a mixture of Bino and Higgsino, satisfying the thermal dark matter relic density, is investigated. 
In such a scenario, the lightest top squark decays mostly into $(i)$ a top quark plus the second or third lightest neutralino, and $(ii)$ a bottom quark plus the lightest chargino, instead of a decay scenario of the lightest top squark   into a top quark and the  lightest neutralino. 
Final states with $\geq 2$ jets, dileptons, and missing energy are expected in a subsequent decay of the second or third lightest neutralinos into the lightest neutralino via an intermediate slepton (``light sleptons" case) or $Z$ boson (``heavy sleptons" case).
The opposite-sign same flavor dilepton mass distribution after subtracting the opposite-sign different flavor distribution shows a clear edge in the case of  light sleptons. The significance for discovering such a scenario is calculated with optimized cuts in both light and heavy sleptons cases.

\end{abstract}
MIFPA-13-05
\maketitle


 
 

\section{Introduction}

Results from the 8-TeV LHC (LHC8) have put bounds on the masses of the gluino ($\tilde{g}$) and first two generation squarks ($\tilde{q}$) in models of supersymmetry (SUSY). If they have comparable masses, the exclusion limits reach to approximately $1.5$ TeV at $95\%$ CL with $13$ fb$^{-1}$ of integrated luminosity \cite{:2012rz, Aad:2012hm, :2012mfa}.
 The third-generation squarks have also actively  been searched, putting experimental bounds on parameter space in certain decay modes  \cite{ATLASStop1, ATLASStop2}.

Light top squark searches so far have been carried out on the case where the  lightest neutralino ($\neu{1}$) is mainly a Bino and the second lightest neutralino ($\neu{2}$) mainly a Wino. 
In such a scenario, the lightest top squark ($\tilde{t}_1$) decays to $\neu{1}$ and a top ($t$) quark at a branching fraction (${\cal B}$) of nearly $100\%$.
In a recent paper by some of the current authors \cite{Dutta:2012kx}, 
the trijet invariant mass $M3$ was used to reconstruct top quarks in fully hadronic final state of
events with at least four non-$b$ jets, at least two $b$ jets and large missing energy.
There have been other approaches to probe this decay  by different groups \cite{Plehn:2011tg}. 
Reference  \cite{Graesser:2012qy} has studied the  $\tilde{t}_1$  decay in the scenario  where $\neu{1}$  and 
the lightest chargino ($\tilde\chi^\pm_{1}$) are purely Higgsino. 

Due to small electroweak (EW) production the bounds on the neutralinos and charginos are much weaker. 
This sector, along with the sleptons, plays a crucial role in the dark matter physics of supersymmetric models. 
In the $R$-parity conserving Minimal Supersymmetric Standard Model (MSSM), $\neu{1}$ is typically the dark matter (DM) candidate. 
If $\neu{1}$ is purely a Bino, its relic density tends to be large since the annihilation cross-section is smaller than the required thermal annhilation rate  $3\times 10^{-26}$ cm$^3$/sec.\footnote{The correct relic density for pure Bino can still be obtained by different methods, such as coannihilation \cite{Griest:1990kh}, slepton exchange for lighter Bino masses, resonance, or non-thermal production without further annihilation \cite{Allahverdi:2012gk}.}. 
One way to obtain the correct relic density is to consider a thermal, well-tempered $\neu{1}$ which is a mixture of Bino and Higgsino \cite{ArkaniHamed:2006mb, Cheung:2012qy}, while having $\neu{2}$ and $\neu{3}$ as primarily Higgsinos.\footnote{Other options include non-thermal Winos or Higgsinos \cite{Dutta:2009uf}, multi-component dark matter \cite{Baer:2012cf}, or well-tempering with Bino and Wino. We note that if $\neu{1}$ is purely a Wino, the annihilation into $W^{+}W^{-}$ final states is in tension with Fermi data for Wino mass below $\sim 250$ GeV \cite{fermiKoushiappas}.}

The purpose of this paper is to utilize $\tilde{t}_1$ decay  to probe the dark matter sector
in a scenario with $\neu{1}$ as a Bino-Higgsino mixture and $\neu{2,3}$ as mainly Higgsinos.
All three are lighter than the lightest top squark, which is in the sub-TeV range. The main theoretical motivation for considering a light top squark as well as light Higgsinos is Naturalness, while the motivation for the presence of a light Bino is to obtain the correct relic density for $\neu{1}$, since if a sub-TeV 
$\neu{1}$ is purely Higgsino, the relic density is too small \cite{Allahverdi:2012wb}. 

In such a scenario, the lightest top squark mainly decays into $t\, \tilde\chi^0_{2,3}$ and $b\, \tilde\chi^\pm_{1}$,
followed by  $\neu{2,3} \rightarrow Z \neu{1}$ or $ll\neu{1}$ (via an intermdiate slepton state) and 
$\tilde\chi^\pm_1 \rightarrow  l\nu\neu{1}$. 
The final state in  $\tilde{t}_1 \tilde{t}_1^{*}$ events has dileptons with jets and missing energy ($\met$). The cases of light slepton  (whose mass is between $\neu{1}$ and $\neu{2}$) and heavy sleptons are considered in events with at least two leptons,  jets and $\met$.

The dilepton final states investigated in this paper can lead to a quite robust $\tilde{t}_1$ search.
The cross section for  $\tilde{t}_1 \tilde{t}_1^{*}$ production is appreciable at the LHC8  for the mass range between $300$ and $700$ GeV. 
It is shown that the SUSY combinatoric and SM backgrounds are reduced by performing a opposite-sign same flavor (OSSF) minus opposite-sign different flavor (OSDF) subtraction. If slepton masses are between $\neu{2}$ and $\neu{1}$, an edge in the dilepton mass distribution could be visible due to higher branching fractions of $\neu{2,3} \rightarrow ll\neu{1}$ decays.  The presence of a $b$-tagged jet in the final state is key to inferring the production of a third-generation squark.

The most dominant background for $\tilde{t}_1 \tilde{t}_1^{*}$ events is $t\bar t$ plus $n$-jets processes.
ATLAS has reported bounds on the $m_{\st}$-$m_{\neu{1}}$ plane  for single lepton, dilepton, and fully hadronic final states using $13$ fb$^{-1}$ of data at $\sqrt{s} = 8$ TeV \cite{ATLASStop1, ATLASStop2}. Bounds have also been reported for the  $\tilde t_1 \rightarrow b \chpm{1}$ decay mode.


It is worthwhile to point out that  for probing  the dark matter sector with mostly-Bino $\neu{1}$ and mostly-Wino $\neu{2}$, vector boson fusion processes are a powerful tool  \cite{Dutta:2012xe}.
This is somewhat complementary to the strategy in this paper.


The rest of the paper is structured as follows. 
In Section \ref{leptonresults}, the results for the ``light slepton case'' $m_{\tilde t_1} > m_{\tilde\chi^{0}_{2}, \tilde\chi^{\pm}_{1}} > m_{\tilde l} > m_{\tilde\chi^0_1}$ are described, while  results for the ``heavy slepton case" $ m_{\tilde l} > m_{\tilde t_1}>m_{\tilde\chi^{0}_{2}, \tilde\chi^{\pm}_{1}} > m_{\tilde\chi^0_1}$ are in Section \ref{leptonresults2}. 
The dark matter relic density for the different scenarios are discussed in Section \ref{darkmatter}, followed by our conclusions in Section \ref{conclusion}.

\section{Scenarios with Light Sleptons}\label{leptonresults}

In this section, scenarios which satisfy the following mass relation are studied: 
\be
m_{\tilde t_1}> m_{\tilde\chi^0_3}, m_{\tilde\chi^0_2},m_{\tilde\chi^\pm_1} >m_{\tilde l}>m_{\tilde\chi^0_1} \,\,. 
\ee
The possible  $\tilde{t}_1$ decay modes are:
\begin{eqnarray}
\tilde t_1& \rightarrow& t\ \tilde\chi^0_1 \\
\tilde t_1& \rightarrow& t\ \tilde\chi^0_2     \rightarrow t\  l^{\mp} \tilde l^{(\ast)\pm} \rightarrow t\ l^{\mp}  l^{\pm}  \tilde\chi^0_1,\\
\tilde t_1&\rightarrow&  b\  \tilde\chi^+_1 \rightarrow b\  l\bar{\nu}\tilde\chi^0_1\,\, {(\rm or}\ b\ q\bar{q}^{\prime} \tilde\chi^0_1)  \\
\tilde t_1& \rightarrow& b\ \tilde\chi^+_2  \rightarrow b\ Z \tilde\chi^+_1 
\end{eqnarray}

\noindent Throughout this paper inclusion of charge conjugate modes is implied. The last mode is allowed when the $\chpm{2}$ is lighter than $\tilde t_1$.

It is clear that one obtains an edge in the dilepton invariant mass distribution as well as $Z$-peak depending on the size of ${\cal B}(\tilde{t}_1 \rightarrow b\, \chpm{2})$ value.
A mass spectrum at our benchmark point is displayed in Table~\ref{benchparameters0}. 

The mass spectrum of the model is determined using \isasugra\ \cite{isajet}. The spectrum is then fed to \pythia\ \cite{pythia}, which generates the Monte Carlo hard scattering events and hadron cascade. These events are passed to the detector simulator \pgs \ \cite{pgs}. We use \darksusy\ \cite{Gondolo:2002tz} for relic density computations.

The latest ATLAS exclusion limits on the chargino-neutralino plane are given in \cite{atlaschpm1}, in events with $3l \, + \, \met$ in $13.0$ fb$^{-1}$ of data at $\sqrt{s}=8$ TeV. Exclusion plots for the direct production of charginos, neutralinos, and sleptons are also given in \cite{atlaschpm2}, in final states with at least two hadronically decaying $\tau$s and $\met$. Our benchmark values of $\neu{1}, \chpm{1}$ are outside the exclusion limits given by ATLAS.

\begin{table}[!htp] 
\caption{SUSY masses (in GeV) at ``light slepton" benchmark point. }
\label{benchparameters0}
\begin{center}
\begin{tabular}{c c c} \hline \hline
 Particle  & Mass (GeV)      & ${\cal B}$  \\ \hline  \\[-.1in]
  $\tilde t_1$  & $500$     &  $17\%$ ($t\neu{2}$), $22\%$ ($t\neu{3}$), $8\%$ ($t\neu{1}$) \\
   &    &                            $53\%$ ($b\tilde\chi^\pm_1$)\\\hline \\[-.1in]
  $\neu{2} $ & $175$       &  $100\%$ ($l\tilde l$)  \\
  $\neu{3} $ & $176$       & $88\%$ ($l\tilde l$)   \\
$\chpm{1} $ &  $164$    & $22\%$ ($l \nu\neu{1}$)\\
  $\tilde l $ & $144$       & $100\%$ ($l\neu{1}$)  \\
  $\neu{1}$  & $113$  & \\    \hline \hline
\end{tabular}
\end{center}
\end{table}

In the benchmark scenario, the mass difference between $\neu{2.3}$ and $\neu{1}$ is around $63$ GeV, and thus an edge in the dilepton invariant mass distribution may be expected around this value.  
The final state of $2$ jets $+$ $2$ leptons $+$ $\met$ events  arises mostly from a combination of  the $\tilde{t}_1 \rightarrow b\tilde{\chi^{\pm}_1}$ and $\tilde{t}_1 \rightarrow t\, \neu{2,3}$  decays. 
If both top squarks decay into a $b$ and a $\chi^{\pm}_1$, then  $2b$ + $2l$ + $\met$ events are expected.

For the light slepton case, results will be shown for $m_{\tilde t_1}$ = $390, 440, 500, 550$ and $600$ GeV (with masses of $\neu{1},\neu{2},\neu{3}$ at the benchmark values in Table \ref{benchparameters0}). The heavy slepton case considered in the next section has $m_{\tilde t_1} = 390$ GeV, with chargino and neutralinos similar to Table \ref{benchparameters0}). 

We note that there is no stringent constraint on the $\tilde{t}_1$ mass from $\tilde{t}_1 \rightarrow  b \chi^{\pm}_1$ mode for the benchmark point in Table \ref{benchparameters0}) with $m_{\tilde t_1} = 500$ GeV. The ATLAS search ~\cite{ATLASStop2} in final states with two leptons places limits upto a stop mass of $\sim \, 450$ GeV in the case where the mass difference between $\tilde{t}_1$ and $\chpm{1} $ is $10$ GeV. The ATLAS search ~\cite{ATLASStop1} in 1 lepton + 4 jets + $\met$ final state places bounds for $\tilde{t}_1$ mass less than $\sim 390$ GeV. 

We note, moreover, that these bounds assume ${\cal B}(\tilde{t}_1 \rightarrow  b \chi^{\pm}_1) = 100\%$  in  $\chi^{\pm}_1 \rightarrow W^{(\ast)}  \neu{1}$, whereas for the benchmark scenario considered in Table \ref{benchparameters3} in the next Section ("heavy slepton" scenario) with $m_{\tilde t_1} = 390$ GeV, the branching fraction to $b \chi^{\pm}_1$ is $\sim 60\%$. Thus, the benchmark point for the "heavy slepton" scenario too is beyond the ATLAS bounds.

The ATLAS top squark searches ~\cite{ATLASStop1} in single lepton+$\geq$4 jet final state assume ${\cal B}(\tilde{t}_1 \rightarrow  t \neu{1}) = 100\%$. For the top squark masses shown in the paper, the  $\tilde t_1$  branching fraction into $t\neu{2,3}$ and  $t\neu{1}$ are $\sim$40$\%$ and $\sim$8$\%$ respectively. Thus, these benchmark points are not ruled out by the relevant ATLAS searches.

ATLAS top squark searches in the fully hadronic final state ~\cite{ATLASStophadronic} also assume ${\cal B}(\tilde{t}_1 \rightarrow  t \neu{1}) = 100\%$ and for the benchmark points in this paper, the exclusion limits are weaker than the limits in the semileptonic final state. Thus, the benchmark points are not ruled out by the fully hadronic searches either.

 
%

%

The analysis begins with selecting events with the following selection cuts:

\begin{enumerate}
\item[($i$)]
At least $2$ isolated  leptons ($e$ or $\mu$) with $\pT  > $  20 and 10  GeV in $|\eta| \, < \, 2.5$, where the isolation is defined as  $\sum p_{T }^{\rm track} \, < \, 5$ GeV with $\Delta R = 0.4$;
\item[($ii$)]
At least $2$ jets with $\pT > 30$ GeV in $|\eta| \, < \, 2.5$;
\item[($iii$)]
At least 1 $b$-tagged jet with $\pT > 30$ GeV in $|\eta| \, < \, 2.5$;
\item[($iv$)] $\met \, > \, 150$ GeV;
\item[($v$)]  $\HT \, > \, 100$ GeV.
\end{enumerate}

At this stage, the dominant SM background is $\ttbar$ events.
OSSF dileptons  arising from the $\neu{2}$ decay  are kinematically correlated and its dilepton invariant mass distribution is expected to have an edge given by
\be
M^{\rm edge}_{ll} \,\, \sim \,\, m_{\neu{2}} - m_{\neu{1}} \,\,.
\ee

The OSSF dilepton mass distribution from $\ttbar$ events can be modelled by the dilepton distribution of OSDF dilepton events \cite{Chatrchyan:2012te}. The OSSF dilepton mass distribution  from supersymmetric combinatoric background ($i.e.$, uncorrelated leptonic pairs) can also be modelled by OSDF dilepton mass distribution. 
This leads to adoping subtracting OSDF distributiuon from OSSF distribution.
The ``light slepton"  benchmark events would arise in an excess in OSSF$-$OSDF dilepton mass distribution.

\begin{figure}[!htp] 
\centering
\includegraphics[width=3.5in]{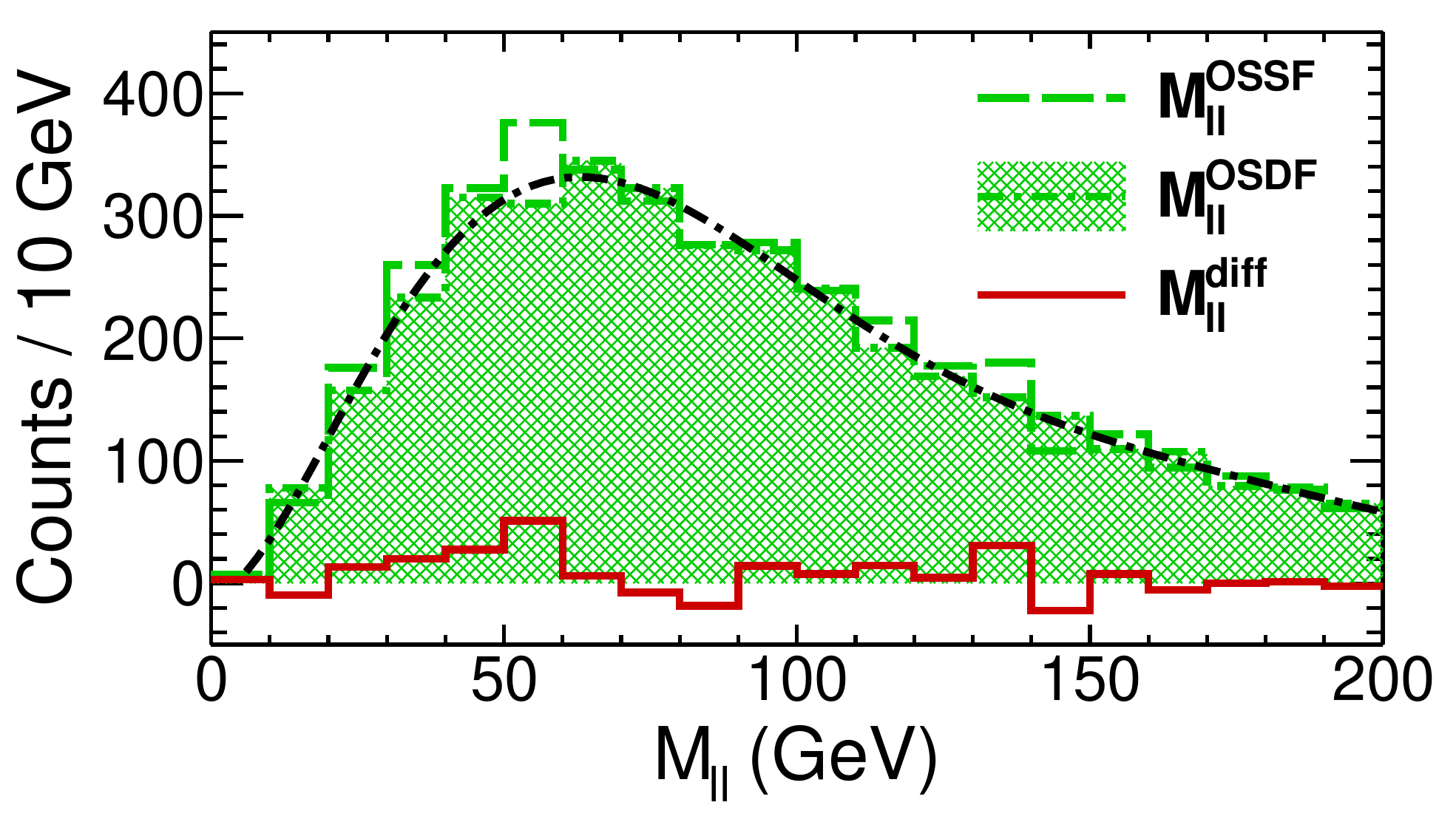}
\caption{The dilepton invariant mass distributions for $t \overline{t} + (0-4)$ jets background and the benchmark point in Table~\ref{benchparameters0} are displayed for  $30$ fb$^{-1}$ luminosity. 
The unshaded histogram shows the $M^{\rm OSSF}_{ll}$ distribution, while the shaded histogram shows the $M^{\rm OSDF}_{ll}$ distribution, which is fitted with the dot-dashed curve. The solid curve shows the subtracted $M^{{\rm diff}}_{ll}$ distribution.}
\label{MllOS_stopChange_susy1a1_lum30_sub_DFfitting_10p}
\end{figure}

 The OSDF dilepton  mass distributions for the SUSY benchmark point in Table \ref{benchparameters0} along with SM $\ttbar$ + (0-4)~jets background is shown in shaded histogram in Fig.~\ref{MllOS_stopChange_susy1a1_lum30_sub_DFfitting_10p}, while its OSSF distribution (blank histogram)  is overlayed. A clear edge is seen at around 63 GeV for  30 fb$^{-1}$ luminosity.

\begin{figure}[!htp] 
\centering
\includegraphics[width=3.5in]{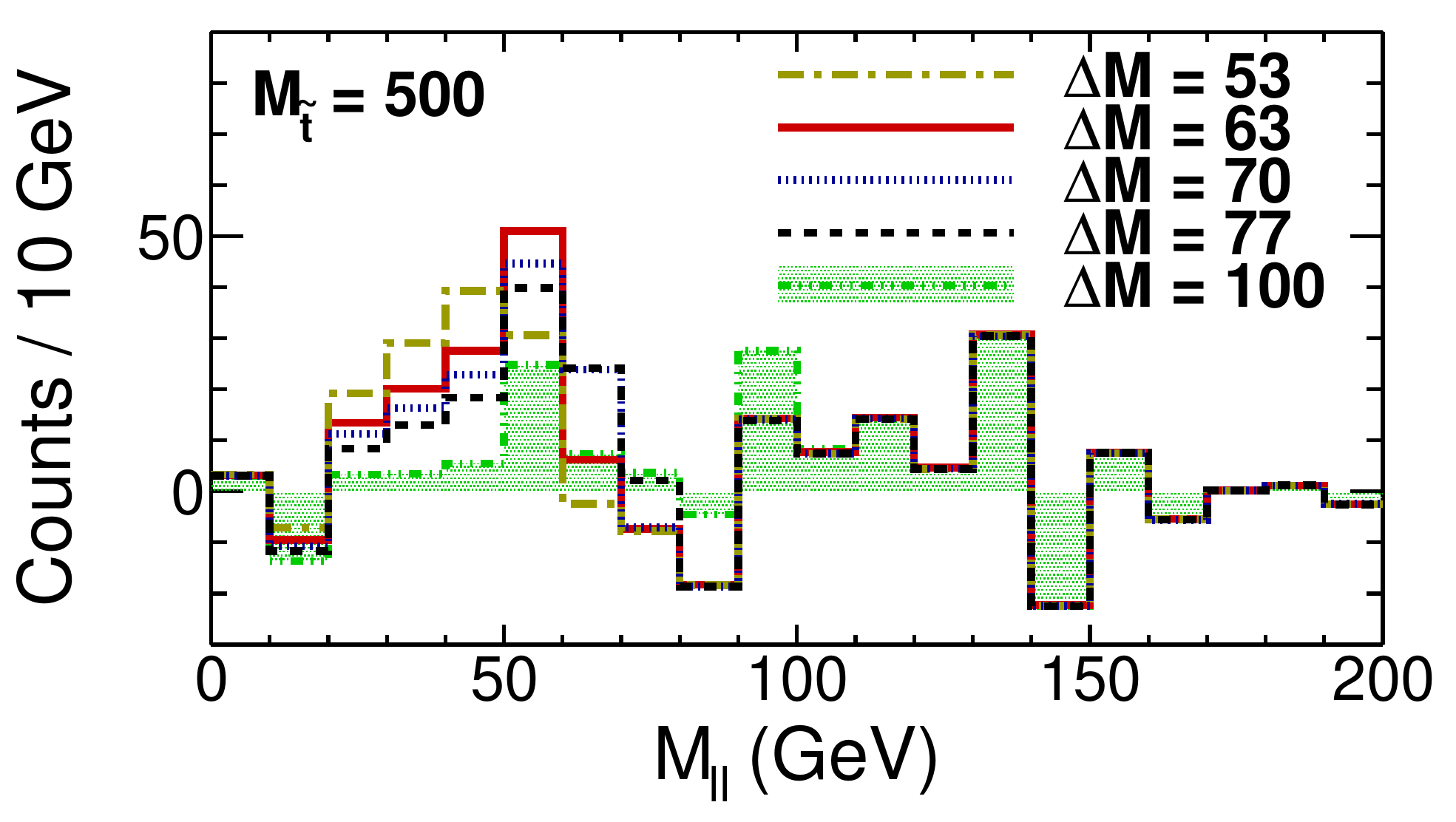}
\caption{The subtracted dilepton invariant mass distribution $M^{{\rm diff}}_{ll}$ as $\Delta M = M_{\neu{2}} - M_{\neu{1}}$ is varied, for $\tilde{t}_1 = 500$ GeV and $\neu{1} = 113$ GeV for  $30$ fb$^{-1}$ luminosity.}
\label{MllOS_edgeShift_susyabcttjf_lum30_sub_DFfitting_10p}
\end{figure}

Figure~\ref{MllOS_edgeShift_susyabcttjf_lum30_sub_DFfitting_10p} shows the flavor subtracted distributions at $\Delta M$ = $m_{\neu{2}} -  m_{\neu{1}}$ =  $53, 63, 70, 77$ and $100$ GeV for $\tilde{t}_1 = 500$ GeV,  $\neu{1} = 113$ GeV and
$m_{\neu{3}} \sim m_{\neu{2}}$. The dilepton mass distribution edge for all these mass differences except $\Delta M = 100$ GeV can be seen clearly. For $\Delta M = 100$ GeV, the signal acceptance is lower and the dilepton study does not have sensitivity at $30$ fb$^{-1}$ luminosity.

\begin{figure}[!htp] 
\centering
\includegraphics[width=3.5in]{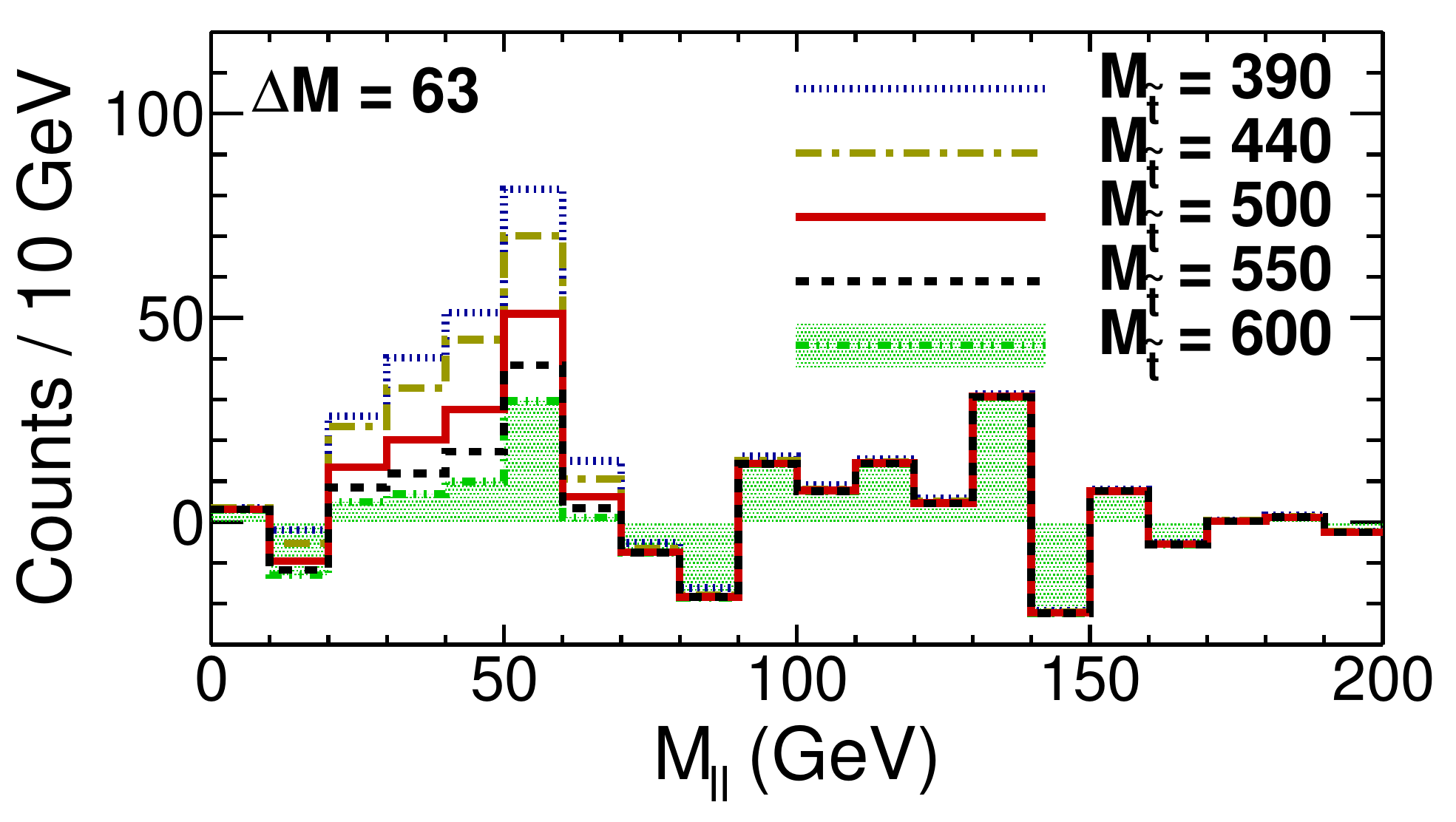}
\caption{The subtracted dilepton invariant mass distribution $M^{{\rm diff}}_{ll}$ as the $\tilde{t}_1$ mass is varied, all other masses remaining at the benchmark value in Table~\ref{benchparameters0} for  $30$ fb$^{-1}$ luminosity.}
\label{MllOS_stopChange_susyabcttjf_lum30_sub_DFfitting_10p}
\end{figure}

Figure~\ref{MllOS_stopChange_susyabcttjf_lum30_sub_DFfitting_10p} shows the flavor subtracted distribution for $\ttbar$ + (0-4)~jets background plus signal events  for $m_{\tilde{t}}$  = $390, 440, 500, 550$ and $600$ GeV, with $m_{\neu{1}} = 113$ GeV and $m_{\neu{3}} \sim m_{\neu{2}}$ = $175$ GeV. The edge of the dilepton distribution for $m_{\tilde{t}}$ mass  upto $550$ GeV can  be distinguished from the background for  $30$ fb$^{-1}$ luminosity.


An excess for each top squark mass case in  Fig. \ref{MllOS_stopChange_susyabcttjf_lum30_sub_DFfitting_10p} is evaluated  in terms of significances (${\cal S}$) in Table \ref{stopsignificance}. We show significances in  cases where  at least one of the jets is required to be a $b$ -jet.
Here ${\cal S} = N_{\rm S} / \sqrt{N_{\rm S} + N_{\rm B}}$, where $N_{\rm S}$ and $N_{\rm B}$ are the number of OSSF dilepton events in range of $20$ GeV  $<  M_{ll}  <$  $70$ GeV  for signal (S) and background (B), respectively.

$N_{\rm B}$ is determined by fitting  the entire (SUSY plus $\ttbar$) OSDF dilepton distribution to a polynomial function and calculating the number of events in 20 GeV  $<  M_{ll}  <$  70 GeV. $N_{\rm S}$ is the the number of events in excess above $N_{\rm B}$. We find that the significance of the benchmark scenario for $m_{\tilde t_1} = 500$ GeV is above 3 $\sigma$ for $30$ fb$^{-1}$

 %
\begin{table}[!htp] 
\caption{[Light slepton case] Signal and background cross sections in fb for various $\tilde{t}_1$  masses  with $\neu{1} = 113$ GeV and $\neu{2},\neu{3} = 175$ GeV. Significances (${\cal S}$) are given at $30$ fb$^{-1}$ }
\label{stopsignificance}
\begin{center}
\begin{tabular}{c  c c  c } \hline \hline \\[-.1in]
   $\tilde{t}_1$  Mass    & Signal & Background  & ${\cal S}$    \\ 
                        (GeV)    & &  & ($\geq 1$ $b$ )     \\ \hline  \\[-.1in]
$390$    & $7.08$   & $46.4$       &  $5.3$  \\
$440$    & $6.0$   & $45.6$    & $4.6$  \\
$500$ (benchmark)     & $3.90$    &  $45.1$    &  $3.1$   \\
$550$    & $2.60$    &  $44.9$       &  $2.1$   \\
$600$    &  $1.70$   &  $44.8$        &  $1.4$   \\ \hline \hline

\end{tabular}
\end{center}
\end{table}

\section{Scenarios with Heavy Sleptons} \label{leptonresults2}

In this section the case where the sleptons are heavier than the lightest top squark in the following relations:

\be
m_{\tilde t_1}>m_{\tilde\chi^0_2},m_{\tilde\chi^\pm_1} >m_{\tilde\chi^0_1} 
\ee

\noindent The possible  $\tilde{t}_1$ decay modes are:
\begin{eqnarray}
\tilde t_1&\rightarrow&  t\,  \tilde\chi^0_2 \rightarrow t\, l l \tilde\chi^0_1\,\, {(\rm or}\ t\, q\bar{q} \tilde\chi^0_1) \\ 
\tilde t_1&\rightarrow&  b\,  \tilde\chi^\pm_1\rightarrow b\,  l\bar{\nu}\tilde\chi^0_1\,\, {(\rm or}\ b\, q\bar{q}^{\prime} \tilde\chi^0_1)  \\
\tilde t_1&\rightarrow&  t\  \tilde\chi^0_1
\end{eqnarray}

\noindent The leptons and quarks  from the $\neu{2}$ and $\tilde\chi^{\pm}_1$ decays are through off-shell $Z$ and $W$ bosons.
SUSY masses at a benchmark point are shown in Table~\ref{benchparameters3}.
\begin{table}[!htp] 
\caption{SUSY masses (in GeV) at ``heavy slepton'' benchmark point.   }
\label{benchparameters3}
\begin{center}
\begin{tabular}{c c c} \hline \hline
Particle & Mass     & ${\cal B}$  \\ \hline  \\[-.1in]
 
  $\tilde t_1$  & $390$  & $17\%$ ($t\neu{2}$), $14\%$ ($t\neu{3}$) \\
     &   & $7\%$ ($t\neu{1}$), $62\%$ ($b\tilde\chi^\pm_1$)  \\ \hline  \\[-.1in]
  $\neu{2} $ & $174$       &  $7\%$ ($ll\neu{1}$)  \\
  $\neu{3} $ & $175$       & $7\%$  ($ll\neu{1}$)   \\
 $\chpm{1} $ & $164$    & $22\%$ ($l \nu \neu{1} $)  \\
  $\neu{1} $  &$112$  &   \\  \hline \hline
\end{tabular}
\end{center}
\end{table}

The same final state like in the ``light slepton'' scenario, i.e., $2$ jets $+$ $2$ leptons $+$ $\met$  is still the key in this scenario.
Therefore the same event selection with re-optimized cuts on the $\met$ and $\HT$ variables. 

\begin{enumerate}
\item[$(i)$] At least 2 isolated  leptons with $\pT >$ 20 and 10  GeV in $|\eta|  <  2.5$;

\item[$(ii)$] At least 2 jets with $\pT >$ 30 GeV in $|\eta|  <  2.5$;
\item[$(iii)$] At least one $b$-tagged jet with $\pT >$ 30 GeV in $|\eta|  <  2.5$;
\item[$(iv)$]  $\met  >  190$ GeV;

\item[$(v)$] $\HT  >  180$ GeV;

\item[$(vi)$] $20  <  M_{ll} <  70$ GeV.
\end{enumerate}

In Table~\ref{heavyslepresults}, we list our signal and background at different stages of cuts and flavor subtraction. The final significance at $30$ fb$^{-1}$ is $0.97$ if a $b$ jet is required in the event sample. Small value of ${\cal B}(Z \rightarrow ll)$ causes smaller significance  in the ``heavy slepton" case compared to the ``light slepton'' case. 
\begin{table}[!htp] 
\caption{[Heavy slepton case]  Cross section (fb) for signal and background at different stages of event selection   and flavor subtractions are shown for the benchmark point in Table \ref{benchparameters3}. }
\label{heavyslepresults}
\begin{center}
\begin{tabular}{c c c } \hline \hline \\[-0.1in]
Event Selection   & $\tilde{t}_1 \tilde{t}_1^{*}$  & $t{\bar t}$ + jets  \\  \hline  \\[-0.1in]
$N_{l} \geq 2$, $N_{j} \geq 2$, $N_{b}  \geq 1$,   & $2.1$ &  $84.7$  \\
 $\met > 190$ GeV, $H_T > 180$ GeV   &                 &    \\ \hline \\[-0.1in]
OSSF dileptons with    & $0.70$    &    $13.2$    \\
$20<M^{{\rm OSSF}}_{ll}<70$ GeV  & &  \\ \hline \\[-0.1in]
OSDF dileptons with    & $0.44$    &    $12.8$    \\
$20<M^{{\rm OSDF}}_{ll}<70$ GeV & &  \\ \hline \\[-0.1in]
OSSF $-$ OSDF dilepton with       &$0.26$     & $0.40$     \\
$20<M^{{\rm OSSF}-{\rm OSDF}}_{ll}<70$ GeV  & &  \\ \hline\hline
\end{tabular}
\end{center}
\end{table}

\section{Dark Matter Relic Density} \label{darkmatter}




For the spectrum (light slepton case) in Table \ref{benchparameters0}, $\neu{1}$ is a mixture of Bino and Higgsino. In this case, one has $m_{\neu{2}} \sim m_{\neu{3}}$ and this proximity of masses increases the branching ratio to dileptonic final states. The correct relic density may be obtained by choosing the relative amounts of Bino and Higgsino in $\neu{1}$ correctly.  This choice depends on the mass splitting between $\neu{2,3}$ and $\neu{1}$ and hence the position of the edge in the dilepton invariant mass distribution. For the benchmark point of Table \ref{benchparameters0} , $\neu{1}$ is $72\%$ Bino and $28\%$ Higgsino, and the relic density is obtained as $\Omega h^2 = 0.11$, as displayed in the first row of Table \ref{Significancetable}. The significance of the analysis is also given.

We note that constraints on DM direct detection for Bino-Higgsino dark matter depends on the sign of $\mu$, as shown for example in \cite{Cheung:2012qy}. The spectrum considered in this paper may be obtained for either sign of $\mu$, and in fact, for $\mu < 0$, the DM masses considered in the paper are not constrained by current experiments. We note that for $\mu > 0$, the mixed Bino-Higgsino DM in the mass range considered here is ruled out by XENON100.

If the mass splitting between $\neu{2,3}$ and $\neu{1}$ is increased, the relic density drops due to smaller Higgsino component in $\neu{1}$. The correct relic density may be obtained in the light slepton case by coannihilation effects, as displayed in the second row of Table \ref{Significancetable}. Due to a low $\pT$  lepton, the significance for this case is smaller. 

For the spectrum (heavy slepton case) in Table \ref{benchparameters3}, $\neu{1}$ is again a mixture of Bino and Higgsino, but due to heavier slepton, the annihilation of $\neu{1}$ through $t$-channel slepton exchange is not open. Consequently, a larger Higgsino component in $\neu{1}$ is required to fulfil the relic density, and hence the mass splitting between $\neu{2,3}$ and $\neu{1}$ is smaller than the light slepton case. The significance  is presented in the third row of Table \ref{Significancetable}.


%
\begin{table}[!htp] 
\caption{Significances (${\cal S}$), relic density ($\Omega h^2$), and Bino-Higgsino composition in three representing SUSY benchmaark points for 30 fb$^{-1}$ luminosity. 
$\Delta M$ denotes the mass splitting between $\neu{2}$ and $\neu{1}$. For all points, $m_{\neu{1}} = 113$ GeV. }
\label{Significancetable}
\begin{center}
\begin{tabular}{c c c c c } \hline \hline  \\[-.1in]
 Masses  & $\{B,H\}$   & $\Omega h^2$ &  ${\cal S}$  & Comments        \\
 (GeV)  & ($\%$)   &    &     &        \\ \hline  \\[-.1in]

$\Delta M = 64$        	& $\{ 72, 28 \}$ & $0.11$ & $3.1$ & Bino-Higgsino DM  \\
$m_{\tilde{l}} = 157$   	&          &        &    & (Light slepton, Table \ref{benchparameters0}) \\ 
$m_{\tilde{t}_1} = 500$	& & & & \\ \hline \\[-.1in]

$\Delta M = 160$        	& $\{ 96, 4 \}$ & $0.11$ & $0.44$ & Mainly Bino DM  \\
$m_{\tilde{l}} = 123$  	&           &        &   & (Coannihilation) \\ 
$m_{\tilde{t}_1} = 500$ 	& & & & \\ \hline \\[-.1in]

$\Delta M = 62$         	& $\{ 67, 33 \}$ & $0.11$ & $0.97$ & Bino-Higgsino DM  \\
$m_{\tilde{l}} = 4000$ 	&          &        &   & (Heavy slepton, Table \ref{benchparameters3}) \\ 
$m_{\tilde{t}_1} = 390$ 	& & & & \\  \hline \hline

\end{tabular}
\end{center}
\end{table}

If $\neu{1}$ is totally a Bino, $\neu{2}$ a Wino, and the Higgsinos are heavy, the relic density tends to be  high unless the sleptons are light, in which case $\neu{1}$ annihilates through $t$-channel slepton exchange ($\neu{1}$ also need to be less than 100 GeV). However, in this case the branching fraction of $\tilde{t}$ to $\neu{2}$ is quite low, and a very large luminosity is required for this analysis. 

Table \ref{Significancetable} is a summary of  the significances  for three  DM scenarios  in this study. We see that the signal $\geq 2$ jets + 2 leptons + $\met$ with OSSF$-$OSDF shows larger significance when there is a slepton between $\neu{2}$ and $\neu{1}$ which is not unnatural in SUSY models..

\section{Conclusion}\label{conclusion}

In this paper, we show that top squark searches at the LHC8 in dilepton final state could be a powerful probe of the composition of $\neu{1}$ if $m_{\tilde{t}_{1}} \lsim$ 600 GeV. Motivated by Naturalness and the relic density constraint, the specific case of a well-tempered neutralino that is a mixture of Bino and Higgsino is probed using the decay of top squarks. 
A typical benchmark point has $m_{\neu{1}} \sim 113$ GeV, $m_{\neu{2}} \sim m_{\neu{3}} \sim 175$ GeV, and $m_{\tilde{t}} \sim 500$ GeV. 



In such scenarios, the lightest  top squark decays prdominantly into $t\,  \neu{2,3}$ and $b\,  \chi^{\pm}_1$.
Both $\neu{2}$ and $\neu{3}$  decay to  $ll\, \neu{1}$ (via an intermdiate slepton state) or $Z\,  \neu{1}$,
while  $\tilde\chi^\pm_1$ decays into $l\nu\neu{1}$ or $W\neu{1}$ . 
Therefore, the final state has dileptons, jets and missing energy with a clear edge in the  OSSF$-$OSDF dilepton mass distribution. 
This is different from  the most studied  scenarios where  $\neu{1}$ and $\neu{2}$ are  mostly Bino and Wino, respectively, and $\tilde t_1$ deacys into  $t\, \neu{1}$ with branching fraction of $100\%$.

In ``light slepton'' scenario where high yield of dilepton events is expected, a discovery sensitivity up to 600 GeV of $m_{\tilde{t}_{1}}$ with 30 fb$^{-1}$ of integrated luminosity at the LHC8 is expected.
If  the Higgsino component in $\neu{1}$ is reduced, then one needs coannihilation processes to satisfy the relic density. 
In such a case, the $\pT$ of leptons becomes lower and the significance of the study is decreased. 
In ``heavy slepton'' case, dileptons are produced from the $Z$ boson decays. Small branching fraction of ${\cal B}(Z \rightarrow ll)$ results in decreasing the discovery sensitivity, compared to the ``light slepton'' case. \\

\section{Acknowledgements}

This work is supported in part by DOE Grant No. DE-FG02-95ER40917 and
by the World Class University (WCU) project through the National Research Foundation (NRF) of Korea funded by the Ministry of Education, Science, and Technology (Grant No. R32-2008-000-20001-0).

\end{document}